\documentclass[conference]{IEEEtran}
\IEEEoverridecommandlockouts
\pdfoutput=1

\usepackage{cite}
\usepackage{amsmath,amssymb,amsfonts}
\usepackage{graphicx}
\usepackage{textcomp}
\usepackage{xcolor}
\usepackage{subcaption}
\usepackage{wrapfig}
\usepackage{multirow}
\usepackage{booktabs}
\usepackage{enumitem}
\usepackage{comment}
\usepackage{soul}
\usepackage{url}
\usepackage{hyperref}
\usepackage{algorithm}
\usepackage{algpseudocode}
\usepackage[T1]{fontenc}

\usepackage{tikz}
\newcommand*\circled[1]{\tikz[baseline=(char.base)]{\node[circle, fill=black, inner sep=0.1ex, text=white] (char) {\scalebox{0.8}{#1}};}}

\def\BibTeX{{\rm B\kern-.05em{\sc i\kern-.025em b}\kern-.08em
    T\kern-.1667em\lower.7ex\hbox{E}\kern-.125emX}}
\makeatletter
\def\ps@IEEEtitlepagestyle{%
  \def\@oddfoot{\mycopyrightnotice}%
  \def\@evenfoot{\mycopyrightnotice}%
}
\def\mycopyrightnotice{%
  \begin{minipage}[t]{\textwidth}\centering\footnotesize
  \copyright~2026 IEEE. Personal use of this material is permitted. Permission from IEEE must be obtained for all other uses, in any current or future\\
  media, including reprinting/republishing this material for advertising or promotional purposes, creating new collective works, for resale or\\
  redistribution to servers or lists, or reuse of any copyrighted component of this work in other works.
  \end{minipage}%
}
\makeatother

\begin{document}
\bstctlcite{IEEEexample:BSTcontrol}

\title{Energy-Efficient Multimodal Inference Serving with Tri-serve}

\author{%
\IEEEauthorblockN{%
Ziyang Jia\textsuperscript{*},
Sara Rashidi Golrouye\textsuperscript{*},
Laxmi Bhuyan\textsuperscript{*},
Benjamin Kubwimana\textsuperscript{\dag},\\
Devashree Tripathy\textsuperscript{\ddag},
Zexin Li\textsuperscript{\S},
Cong Liu\textsuperscript{\S},
Daniel Wong\textsuperscript{\S}}
\IEEEauthorblockA{%
\textsuperscript{*}\textit{Department of Computer Science and Engineering, University of California, Riverside}\\
\textsuperscript{\dag}\textit{NVIDIA} \\
\textsuperscript{\ddag}\textit{Department of Computer Science and Engineering, Indian Institute of Technology, Bhubaneswar}\\
\textsuperscript{\S}\textit{Department of Electrical and Computer Engineering, University of California, Riverside}\\
\texttt{\{zjia016, srash034, lbhuyan\}@ucr.edu, bkubwimana@nvidia.com,}\\
\texttt{devashreetripathy@iitbbs.ac.in, \{zli536, congl, danwong\}@ucr.edu}}
}

\maketitle

\begin{abstract}
Multimodal model inference creates substantial energy demand with growing performance requirements. %
Within GPUs, power is autonomously managed by an on-board power management unit (PMU), which makes frequency boosting/throttling decisions. However, we find that these hardware-managed frequency decisions can cause significant power inefficiency.
This work identifies three classes of power inefficiencies within modern multimodal inference serving: 
(1) inter-stage dependency stalls run at near-maximum frequency despite being idle; (2) anti-correlation between auto-boost frequency and arithmetic intensity (A.I.) results in compute-bound phases (e.g., prefill) running at lower frequency and vice versa; and (3) thermal throttling degrades SM frequency and throughput.

We propose \textbf{Tri-serve}, a software-based DVFS controller that jointly accounts for inter-stage dependency stalls, the arithmetic-intensity effect on frequency and power, and the thermal-throttling effect of high A.I. phases, to deliver energy-efficient multimodal serving on commodity GPUs. We show that Tri-serve achieves a 22\% energy-efficiency improvement with no latency or throughput impact.
\end{abstract}


\section{Introduction}
\label{sec:introduction}

Multimodal model inference now powers production tools such as coding/general agents (Claude Code, Cursor, Codex, OpenClaw, etc.) and conversational assistants (ChatGPT, Gemini, etc.). However, it creates substantial energy demand alongside ever-growing performance requirements. Recent multimodal inference serving stacks, such as vLLM-Omni~\cite{yin2026vllmomni}, have improved job completion time for multimodal inference by fully disaggregating each modality stage across GPUs. However, reducing energy per generated token and power consumption has remained a secondary priority in the rapid evolution of ML systems and infrastructure.

Many prior works have explored GPU Dynamic Voltage-Frequency Scaling (DVFS) for deep-learning inference~\cite{10.1145/3620666.3651329, 10540202, 298496, 10946751,10.1145/3634769.3634808} with the aim of selecting a lower frequency to satisfy service level objectives (SLO).
These efforts primarily target unimodal LLM serving as a black box and select the best frequency level given workload properties, such as batch size, sequence length, etc. 
However, these prior works do not address the unique challenges in multi-stage, multi-modality pipelines that dominate modern multimodal serving, where heterogeneous stages exhibit distinct workload characteristics.

In this work, we explore opportunities for GPU power management for modern multimodal inference serving. %
By profiling vLLM-Omni serving Qwen-Omni~\cite{hui2024qwen2, yang2025qwen3, xu2025qwen3} on commodity datacenter GPUs, we reveal \textbf{three classes of power inefficiency} that are inherent to the GPU's hardware power management unit (PMU) and its management of DVFS auto-boosting capabilities.

\noindent\textbf{Inefficiency 1) Inter-stage dependency stalls waste idle power.}
Qwen-Omni adopts a multi-stage architecture consisting of a Thinker, Talker, and Vocoder model. %
Due to memory limits, these stages are typically mapped onto separate GPUs, forming a producer-consumer pipeline in which downstream stages routinely stall and wait on upstream stages. During these dependency stalls, we observed that the core and memory clocks remain pinned near the auto-boost ceiling, drawing unnecessary power while performing no useful work.

\noindent\textbf{Inefficiency 2) Auto-boost frequency is anti-correlated with arithmetic intensity.}
We observe that compute-bound prefill consistently runs at a \emph{lower} core frequency than memory-bound decode and idle dependency stall phases. That is, the GPU's frequency levels are typically lower when more performance is necessary (higher arithmetic intensity), and frequency is higher when performance is not necessary (lower arithmetic intensity).

\noindent\textbf{Inefficiency 3) Thermal throttling causes frequency degradation in compute-intensive phases. }
We observe that the prefill phases of the Thinker and Talker stages suffer from a monotonic decay of core frequency over time. We identify that this is due to thermal throttling of the SMs (GPU cores), where compute-heavy phases run hotter. The GPU PMU manages auto-boost frequencies based on the power and thermal headroom that is available~\cite{10.1145/1815961.1815998,guerreiro2018gpgpu,5450539}.
Therefore, compute-heavy phases tend to deplete the thermal headroom the longer they run, resulting in the frequency decrease over time. We observe that this thermal throttling phenomenon is universal across multiple GPU models.

Based on these observations, we overcome the limitations of hardware PMU-managed frequency scaling by proposing \textbf{Tri-serve}, a software-based end-to-end DVFS controller that jointly accounts for inter-stage dependency stalls, the arithmetic-intensity effect on frequency and power, and the thermal-throttling effect of the PMU, to deliver energy-efficient multimodal serving. Our contributions are:
\begin{itemize}[nosep,leftmargin=*]
    \item We perform a characterization of dependency-induced idle power inefficiency in multimodal inference pipelines. To solve this,  we propose \textit{stall-aware idle frequency scaling} that minimizes both core and memory clocks whenever a GPU is blocked due to inter-stage dependency stalls.
    \item We introduce a \textit{frequency-locked roofline microbenchmark} that enables us to characterize the behavior of the GPU's PMU frequency management policy. We found that workload arithmetic intensity and SM frequency are anti-correlated, resulting in high frequency when performance is not necessary, and vice versa. To solve this, we propose an \textit{arithmetic intensity-aware frequency scaling} policy that selects an optimal frequency given the A.I. %
    \item We observe strong coupling between the die temperature, core frequency, and TFLOPS. We observe that this thermal throttling phenomenon is ubiquitous across many GPUs. The magnitude of the losses is amplified with higher A.I., compounding the aforementioned anti-correlation effect. To overcome this, we introduce a  \textit{thermal-aware frequency scaling} policy that identifies sustainable frequencies to avoid thermal throttling of compute-heavy phases.
    \item Finally, we introduce Tri-serve, a software-based DVFS solution for inference serving of multimodal models that integrates the aforementioned three optimizations. Through a real implementation on a GPU cluster, we show that Tri-serve can achieve a $\sim$20\% improvement in energy per output token without impacting latency or throughput.
\end{itemize}

\section{Characterizing Energy Inefficiencies in Multimodal Serving}
\label{sec:motivation}

\subsection{Multimodal Model Architecture Background}
\label{sec:motivation-pipeline}

Multimodal Large Language Models (MLLMs), such as the Qwen-Omni
series~\cite{xu2025qwen3,team2026qwen3}, extend generative AI to text, image,
and audio inputs and outputs~\cite{xie2025mme}. Unlike text-based LLMs, MLLMs
adopt \emph{pipelined} architectures whose stages handle modality-specific
encoding, reasoning, and generation in sequence.
vLLM-Omni~\cite{yin2026vllmomni}, a representative high-throughput
multimodal serving framework, executes each stage as an independent worker
process on top of PagedAttention~\cite{10.1145/3600006.3613165} memory
management. In this work, we use Qwen-Omni as the representative multimodal model.

\begin{figure}[!t]
  \centering
  \includegraphics[width=\linewidth]{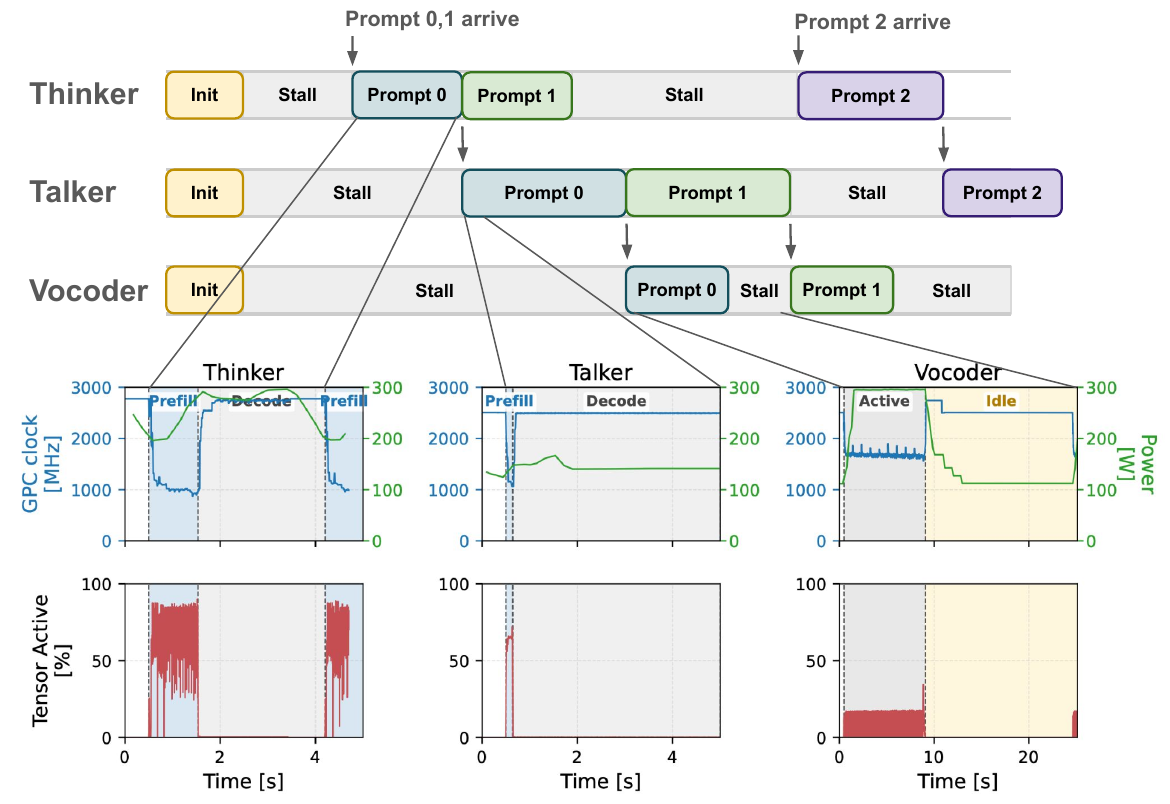}
  \vspace{-5mm}
  \caption{Pipeline of the three Qwen-Omni stages and their SM frequency, power, and tensor activity profiles.}
  \label{fig:vllm_omni_pipeline}
  \label{fig:qwen_5prompt_nsysprofileview}
  \vspace{-3mm}
\end{figure}

The Qwen-Omni pipeline is disaggregated into three distinct stages: Thinker, Talker, and Code2Wav/Vocoder.
The first stage, Thinker, 
is the core multimodal LLM
    responsible for reasoning. Front-end vision and audio encoders convert
    image/video and waveform inputs into a token stream that, together with the
    text prompt, is consumed by the Thinker in a compute-bound \emph{prefill}
    phase. Thinker then emits text tokens and per-token hidden states
    in an autoregressive \emph{decode} loop. 
    
    The next stage, Talker, is another autoregressive transformer
    that consumes the Thinker hidden states one at a time and emits discrete
    acoustic tokens (RVQ codes). Each Thinker hidden state triggers a
    short Talker step, so the Talker inherits the prefill--decode
    structure of unimodal LLMs.
    The last stage, Code2Wav/Vocoder, is a non-autoregressive
    synthesis model (Diffusion Transformer/DiT~+~codec) %
    that converts batches of Talker acoustic codes into the final audio waveform in chunked, one-shot bursts. It activates sparingly only when the Talker has accumulated enough
    tokens to fill a chunk.

\noindent\textbf{Pipeline parallelism: } Figure~\ref{fig:vllm_omni_pipeline} shows an illustrative example where each stage in Qwen-Omni maps onto an individual GPU\footnote{Note that multiple stages can also be mapped to the same GPU, which we also explore in Section~\ref{sec:baseline_comparison} with Thinker and Vocoder sharing a GPU.}.  vLLM-Omni organizes these stages in a \emph{pipeline-parallel}
configuration (one stage per device) and connects adjacent
stages with a Python producer/consumer queue. %

The mapping is dictated by
the stages' divergent memory footprints (each stage holds its own model
weights and KV cache, which together exceed a single 48~GB device) with 
overlapping compute. For example, Thinker decode for request~$r$
runs concurrently with Talker decode for request~$r{-}1$ and Vocoder
synthesis for request~$r{-}2$. Data is exchanged between stages through queues, not 
tensor-parallel collectives.%

\subsection{Inter-Stage Dependency Stalls Waste Idle Power}
\label{sec:motivation-stalls}

To explore the inefficiencies in multimodal serving, we profile the execution of vLLM-Omni using Nsight Systems on NVIDIA RTX 6000 Ada GPUs.
See Section~\ref{subsec:system_setup} for more details on our experimental setup.

\begin{figure}
  \centering
  \includegraphics[width=0.6\linewidth]{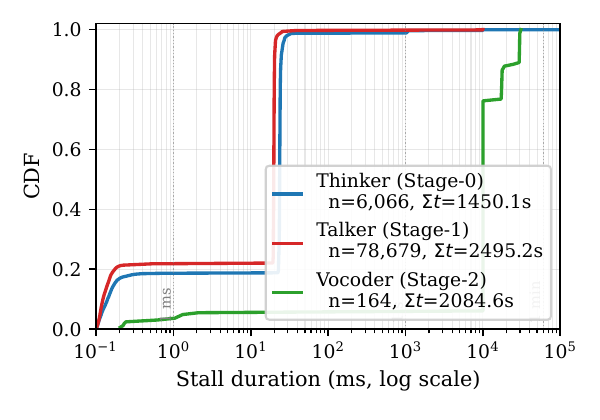}
  \vspace{-5mm}
  \caption{Per-stage CDF of \texttt{sem\_wait} stall durations.}
  \label{fig:stall_cdf}
  \vspace{-3mm}
\end{figure}

The disaggregated Thinker--Talker--Vocoder pipeline introduces a producer-consumer
queue between every pair of stages, as shown in Figure~\ref{fig:vllm_omni_pipeline} (top). 
As autoregressive modules, both Thinker and Talker consist of prefill and decode phases, while Vocoder is a diffusion-based, non-autoregressive synthesis stage.
Since the Thinker waits on incoming requests and each individual stage exhibits different execution times, these inter-stage dependencies result in significant GPU stalls.
Whenever a downstream stage drains its
queue and waits for the next request, it blocks on a \texttt{sem\_wait()} call and the SMs (GPU cores) go idle. 

Figure~\ref{fig:stall_cdf}
reports the CDF of \texttt{sem\_wait()} durations extracted from
the same Nsight Systems trace from the bottom of Figure~\ref{fig:qwen_5prompt_nsysprofileview}.
In a concurrent online serving scenario,
across the three stages, total stall time accounts for roughly
$16\%$ of total GPU time, while the Vocoder suffers from $34\%$ stall time.

In addition, these stall periods are fairly long, typically at least 20\,ms for the Thinker and Talker and above 10\,s for the Vocoder, as shown in Figure~\ref{fig:stall_cdf}.
Although continuous batching allows concurrent requests to interleave, the inherent execution time disparity between the slower autoregressive upstream stages and the faster non-autoregressive downstream stages causes downstream queues to rapidly drain, making these dependency stalls impossible to fully hide.

\begin{figure}

    \centering
    \includegraphics[width=0.7\linewidth]{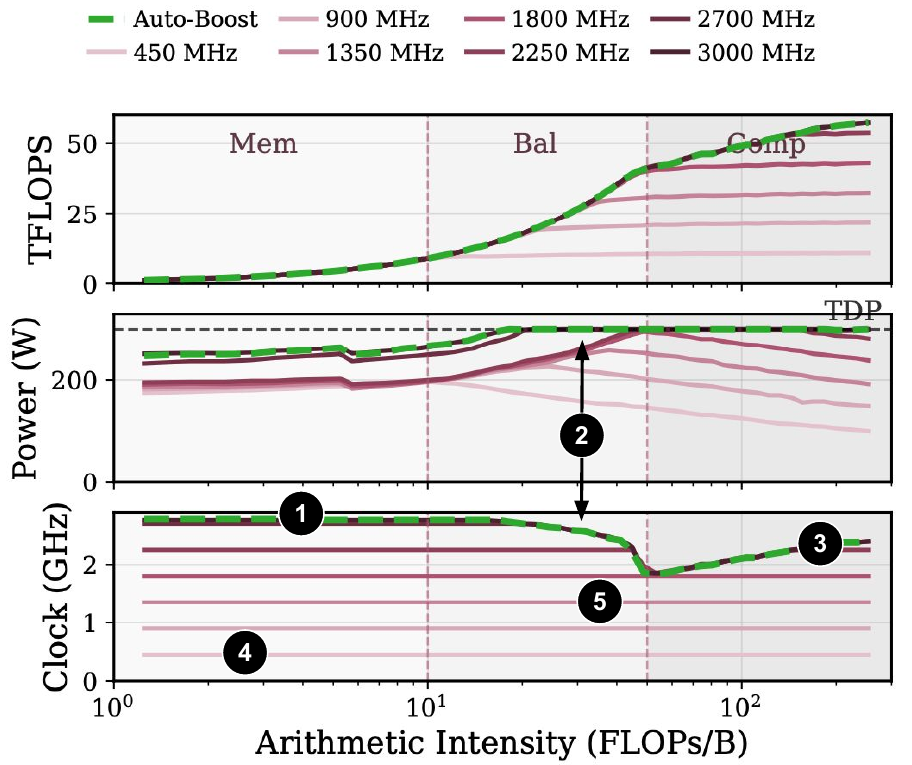}
    \caption{Frequency-locked roofline benchmark results.
    }
    \label{fig:frequency_locked_roofline}
    \vspace{-1em}
\end{figure}

\noindent\textbf{Observation 1: Inter-stage dependency leads to long-running stalls with high SM frequency.}
The bottom of Figure~\ref{fig:qwen_5prompt_nsysprofileview} shows the compute clock frequency, power, and corresponding tensor utilization of the GPU at different phases of each stage, as measured by Nsight Systems.
Looking at the idle period of the GPU load with Vocoder,
the GPU does \textit{not} lower frequency during these idle periods. 
The SM clock stays pinned at $\sim$2500\,MHz, and the per-GPU package power remains above 60\,W even when the SMs are inactive. The frequency at idle is far above the frequency of the P8 idle state (210\,MHz), causing the active-idle power to be higher than the static power; thus, there is significant potential for saving power during these stall periods.
Prior work~\cite{lei2026energy} also observes these inefficient frequencies during ``execution-idle'' phases, validating our findings.

\subsection{PMU-managed Frequency Decisions Are Not Optimized for Varying Arithmetic Intensity}
\label{sec:motivation-ai}

\noindent\textbf{Observation 2: High tensor activity coincides with lower SM
frequency, and vice versa.} 
From the bottom of Figure~\ref{fig:qwen_5prompt_nsysprofileview}, we observe that for Thinker, during the prefill phase (with high-arithmetic-intensity
dense GEMM kernels), the SM frequency \emph{decreases} to $\sim$1000--1200\,MHz, and during the decode phase
(with low A.I. GEMV-like operators) the SM frequency \emph{increases} to the auto-boost ceiling. Essentially, the SM clock frequency is highest when compute performance is unnecessary, and frequency is lowest when compute performance is critical. 
We observe this anti-correlation pattern across both
autoregressive stages in Thinker and Talker. This observation is also consistent with prior observations during ML training collective communication phases, which have low A.I.~\cite{jia2024pccl}.

\subsubsection{Frequency-locked Roofline Benchmarking}  
To explore \textit{why} the PMU exhibits this anti-correlated frequency policy, we introduce a \textit{frequency-locked roofline} benchmark that aims to demystify the PMU's auto-boost arithmetic intensity-dependent behavior. We sweep a synthetic
roofline kernel (based on matrix multiply) across A.I.~$\in\{1,25,50,100,200\}$\,FLOPs/B and lock the SM clock at $f\in\{450,\dots,3000\}$\,MHz, 
comparing against the PMU-managed auto-boost reference. Figure~\ref{fig:frequency_locked_roofline} shows the throughput, power, and core clock across a range of A.I.%

\begin{figure}[t!]
    \centering
    \begin{minipage}{0.5\textwidth}
        \centering
        \includegraphics[width=\linewidth]{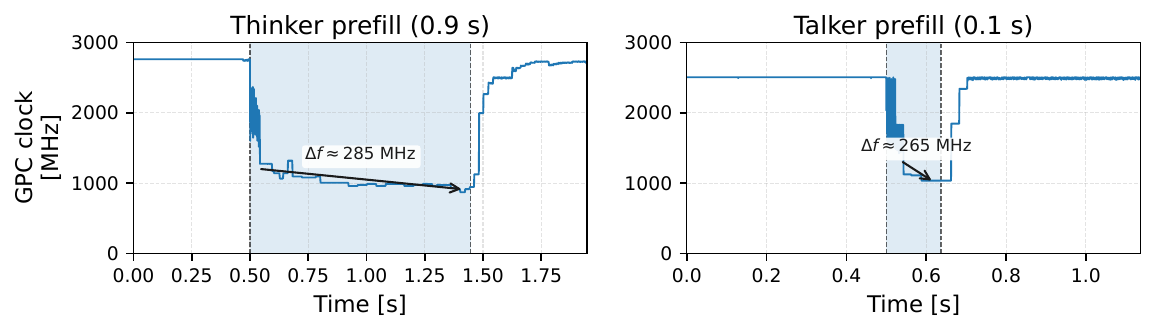}
        \subcaption{Zoom-in on a single Thinker/Talker prefill window. %
        }
        \label{fig:prefill_frequency_drop_zoomin}
    \end{minipage}\hfill
    \begin{minipage}{0.25\textwidth}
        \centering
        \includegraphics[width=\linewidth]{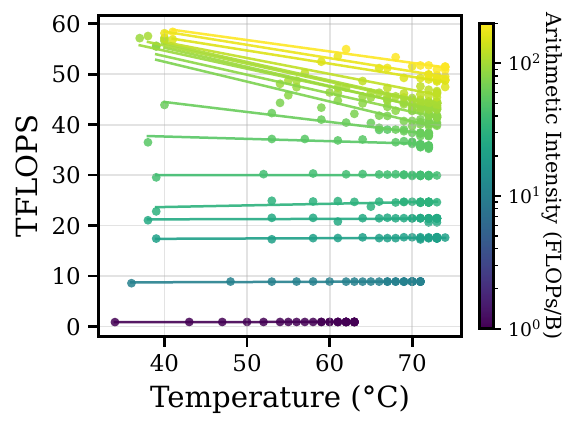}
        \subcaption{TFLOPS vs.\ temperature, colored by A.I. %
        }
        \label{fig:thermal_tflops_vs_temp}
    \end{minipage}\hfill
    \begin{minipage}{0.23\textwidth}
        \centering
        \includegraphics[width=\linewidth]{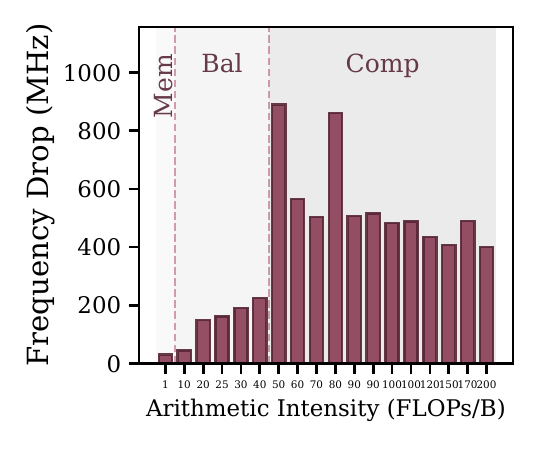}
        \subcaption{Steady-state frequency drop vs.\ A.I. %
        }
        \label{fig:thermal_freq_throttling}
    \end{minipage}
    \caption{A.I.-centric thermal-throttling effect by PMU.}%
    \label{fig:thermal_heatmap_block}
    \vspace{-5mm}
\end{figure}

\textbf{Observation 2a:} \textit{\textbf{The achievable frequency ceiling under PMU-guided 
frequency scaling varies depending on A.I.}}  As shown in Figure~\ref{fig:frequency_locked_roofline} (bottom), A.I. less than 10 Flop/B is considered memory-bound and A.I. greater than 60 Flop/B is considered compute-bound. We observe that during low A.I. periods (i.e., Decode), the PMU's auto-boost (dashed line)
is able to achieve the maximum frequency (\circled{1}). As the A.I. increases towards the ridge, where the workload is compute and memory balanced, the GPU's power hits the maximum observed power level and the core clock gradually begins to decrease (\circled{2})~\cite{10.1109/LCA.2023.3278652}. Then at even higher A.I. levels (i.e., Prefill), the core frequency rises slightly as the workload exhibits less memory activity and reallocates power towards the SMs (\circled{3}).
However, this frequency level is still lower than that achieved at lower A.I. levels (\circled{1}). This is because frequency is bounded by power \textit{and} thermal headroom at higher A.I.

\textbf{Observation 2b:} \textit{\textbf{There exist lower frequencies that achieve the same throughput; thus, auto-boost can waste energy.}} As shown in Figure~\ref{fig:frequency_locked_roofline} (middle, top), at low A.I. the
auto-boost controller selects the maximum frequency (\circled{1}). 
However, that same throughput is achievable with the lowest clock (\circled{4}), thus wasting power.
As A.I. increases, the lowest required frequency to achieve the maximum throughput begins to increase slowly, with the balanced A.I. region requiring 450\,MHz to 1800\,MHz (\circled{5}), still significantly lower than the auto-boost frequency. These results demonstrate the inefficiency of auto-boost and show the necessity for DVFS controllers to be arithmetic-intensity aware to save power.

\subsection{Thermal Throttling Limits Performance During Compute-heavy Phases}
\label{sec:motivation-thermal}

\noindent\textbf{Observation 3: Thermal throttling degrades frequency during high arithmetic intensity phases. } 
Figure~\ref{fig:prefill_frequency_drop_zoomin} shows the frequency behavior during the compute-heavy prefill phases for both Thinker and Talker. Over the course of prefill execution, the clock frequency starts higher (\textasciitilde1300\,MHz), then gradually decreases (down to \textasciitilde1000\,MHz) as prefill progresses, essentially losing \textasciitilde20\% in performance over the course of prefill.
As we will show, this is due to thermal throttling by the PMU as auto-boost frequencies are dependent on the thermal headroom that exists.

\begin{figure}[t!]
    \centering
    \includegraphics[width=\linewidth]{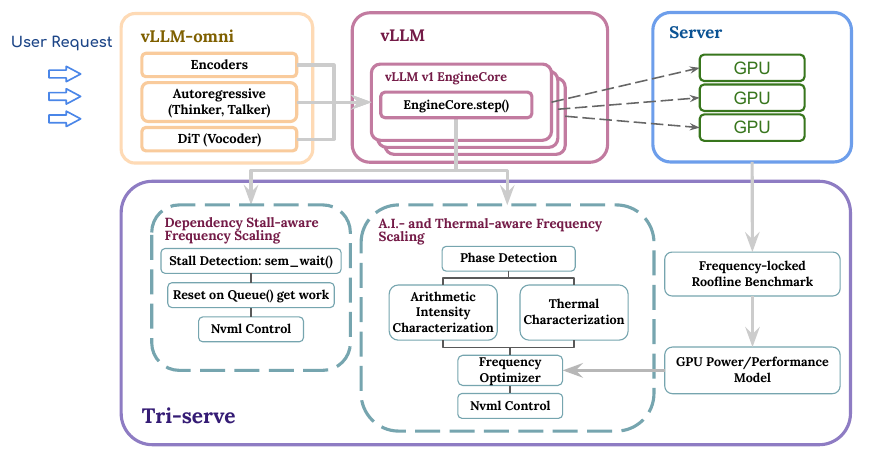}
    \caption{Tri-serve architecture. 
    }
    \label{fig:date_serve_arch}
    \vspace{-3mm}
\end{figure}

\noindent\textbf{Benchmarking Thermal Throttling. } 
To understand how the hardware PMU handles thermal throttling and arithmetic intensity, 
we sweep a range of arithmetic intensity kernels, running each for 240\,s, allowing the GPU to reach a steady-state temperature.

As shown in Figure~\ref{fig:thermal_tflops_vs_temp}, for low A.I. kernels (lower data points), the performance remains relatively stable as the GPU reaches a steady-state temperature. However, as the arithmetic intensity increases (higher data points), there is a clear negative correlation with TFLOPS as the GPU's temperature increases.
Figure~\ref{fig:thermal_freq_throttling} shows the average frequency drop observed across a range of A.I. For compute-heavy kernels, we observe frequency drops of \textasciitilde400--800\,MHz due to thermal throttling.
Compute-heavy kernels tend to generate more heat, resulting in less thermal headroom for the hardware PMU to leverage for auto-boosting, which produces this thermal-throttling effect.

\section{Tri-serve: Dependency-, A.I.-, Thermal-, and Efficiency-aware Management}
\label{sec:system}

We now present \textbf{Tri-serve}, a software-coordinated DVFS controller
for multimodal serving that resolves the PMU-guided auto-boost inefficiency of dependency stalls, anti-correlation of arithmetic intensity and frequency selection, and compute-heavy thermal throttling. Figure~\ref{fig:date_serve_arch} shows an overview of our Tri-serve framework. Tri-serve consists of three main components:
\\
1) \textit{Stall-aware idle frequency scaling:} Resolves dependency stall inefficiencies by minimizing SM/memory frequencies whenever all stages sharing a physical GPU have entered a blocking wait.
\\    
2) \textit{A.I.-aware frequency scaling:} Resolves unnecessarily high frequencies by detecting phase windows (stages, prefill/decode) and selecting the energy-optimal frequency based on the A.I. of that phase with our analytical performance and power models.
\\
3) \textit{Thermal-aware frequency scaling:} Avoids thermal-throttling effects and frequency degradation by initially running at a lower frequency to conserve thermal headroom (pacing) and then running at a higher sustainable frequency to expend that thermal headroom (racing). %

Together, \textit{Stall-aware scaling} and \textit{A.I.-aware scaling} save power during dependency stalls and decode phases, while \textit{Thermal-aware scaling} improves the performance of prefill phases, all working together to improve the energy efficiency of multimodal inference serving.

\subsection{Implementation Details} Each Tri-serve component is designed with a trigger event, a frequency policy, and actuation of frequency change. All three components share a single NVML-backed actuation primitive
(\texttt{nvmlDeviceSetGpuLockedClocks}) to set the desired SM and memory clock. The remainder of this section details the frequency policy for each component.

\subsubsection{Phase Detection}
Trigger events are phases, such as stalls, prefill, and decode. 
Phases are detected through a single phase tag derived from
the vLLM v1 \texttt{EngineCore} scheduler.  %
From the scheduler's per-request \texttt{num\_scheduled\_tokens} plan, a step is labeled
\emph{decode} or \emph{prefill} or \emph{mixed}. In vLLM-Omni, chunked prefill is not supported, so there is no \emph{mixed} phase. Boundaries are anchored by \texttt{torch.cuda.Event} markers injected at the beginning and end of the phases, which we use as trigger events for \textit{A.I.-aware Scaling} and \textit{Thermal-aware Scaling}. Dependency stall trigger events are \texttt{sem\_wait()} calls.

\subsection{Dependency Stall-aware Idle Frequency Scaling}
\label{sec:system-stalls}

This component is triggered by dependency stalls, which occur when a stage blocks at a \texttt{sem\_wait()} call while waiting on a Python queue. On entering the \texttt{sem\_wait()}, we trigger frequency scaling via
\texttt{nvmlDeviceSetGpuLockedClocks()}, and we restore auto-boost on exiting it. During stalls, we set the SM and memory frequency to 210 and 810\,MHz, respectively, which are the frequencies of the lowest-level P8 performance state. Given the coarse-grained nature of dependency stalls (as shown in Figure~\ref{fig:stall_cdf}), this policy presents a low-complexity yet highly effective solution to the power inefficiencies of dependency stalls.

\subsection{Arithmetic Intensity-Aware Frequency Scaling}
\label{sec:system-pd}

The frequency-locked roofline of Figure~\ref{fig:frequency_locked_roofline} shows that
auto-boost is not energy-optimal. The PMU picks high clock frequencies for
memory-bound decode when compute performance is not necessary, and throttles frequency on 
compute-bound prefill when compute performance matters. To address this, \textit{A.I.-aware frequency scaling} aims to first identify the lowest power that achieves the best throughput given a certain arithmetic intensity, then selects the highest frequency that achieves that power level. This Tri-serve component specifically targets decode periods where frequencies are unnecessarily high, and is triggered by instrumentation in the vLLM \texttt{EngineCore} that detects entry into a decode phase.

\label{sec:AI-f-p-modeling}
\noindent\textbf{Modeling A.I.-aware Throughput and Power. }
This component requires both a throughput model, $\Theta(A.I., f)$, to estimate performance given an A.I. and frequency level, and a power model, $P(A.I., f)$, to estimate power trends (Figure~\ref{fig:frequency_locked_roofline}). 

Throughput exhibits a piecewise behavior where the memory-bound component can be modeled separately from the compute-bound component. Therefore $\Theta(A.I.,f)$ is modeled as:
\begin{equation}
  \Theta(A.I., f) = \min(\eta \cdot f, \,\, \beta \cdot A.I.)  
  \label{eq:vanilla_throughput}
\end{equation}
, where the memory bandwidth (memory coefficient) is $\beta$~\cite{williams2009roofline}, so the performance in the memory-bound section is $\beta \cdot A.I.$
and the frequency-capped performance in the compute-bound section is $\eta \cdot f$, where $\eta$ is peak ops per clock.

Power is modeled as: 
\begin{equation}
  P(A.I., f) = \min\left( P_{idle}(f) + P_{dyn}(f) \cdot \Phi(A.I., f), \, P_{TDP} \right)  
  \label{eq:vanilla_power}
\end{equation}

, where $P_{TDP}$ is the maximum power consumption of the GPU, $P_{idle}(f)$ is the active idle power drawn when all SM cores are idle for a given frequency, and $P_{dyn}(f)$ is the dynamic power that scales with the compute and memory utilization factor, $\Phi(A.I., f)$.
The utilization factor is:  
\begin{equation}
\Phi(A.I., f) = \min\left( \frac{\eta \cdot f}{\beta \cdot A.I.} ,  \frac{\beta \cdot A.I.}{\eta \cdot f} \right)    
\end{equation}
Notice in Figure~\ref{fig:frequency_locked_roofline} that the power tends to increase as A.I. increases, until the point where compute and memory are balanced; then the power begins to decrease again (assuming it is not power limited by the TDP). This is because in the memory-bound phase, as A.I. increases, we begin to add more compute activity in addition to the existing memory activity. Power is at the peak where both are balanced because both memory and compute are stressed equally. Then as A.I. increases further into the compute-bound phase, memory activity decreases and compute power dominates, resulting in a drop in the power. $\Phi(A.I., f)$ captures this transition between memory and compute utilization.

\begin{figure}[t]
    \centering
    \begin{minipage}{0.49\linewidth}
        \centering
        \includegraphics[width=\linewidth]{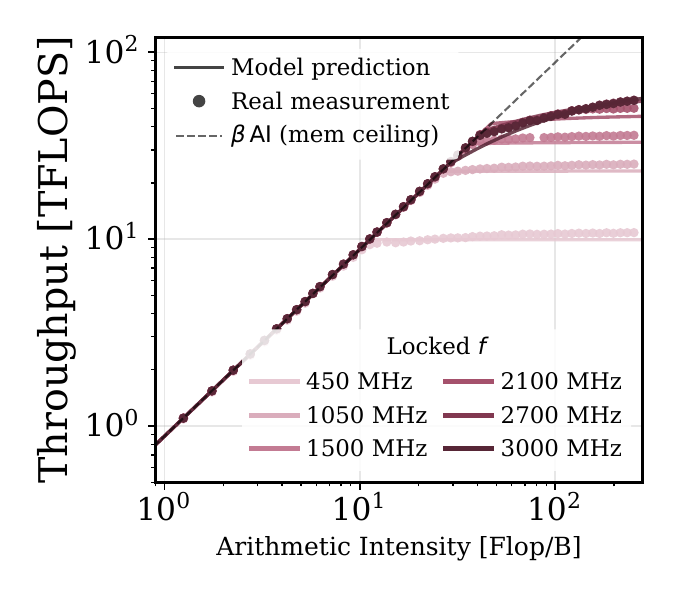}
        \subcaption{Throughput model $\Theta(A.I.,f)$ vs measured.}
        \label{fig:iccd_fit_throughput}
    \end{minipage}\hfill
    \begin{minipage}{0.49\linewidth}
        \centering
        \includegraphics[width=\linewidth]{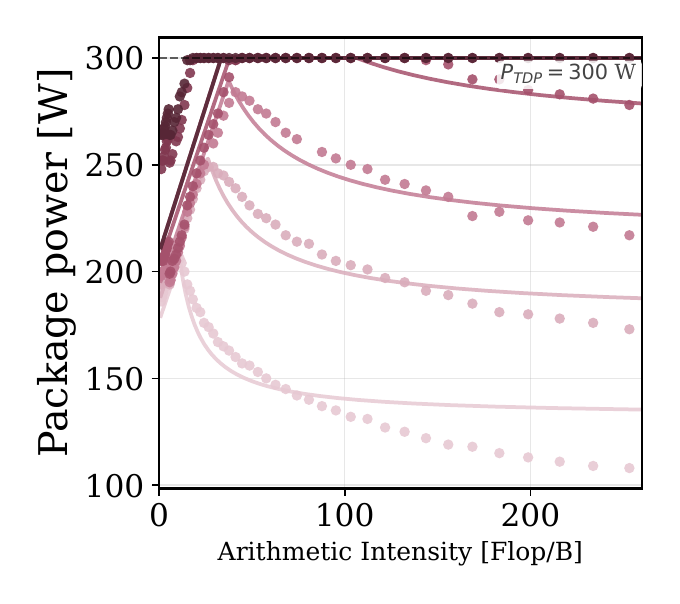}
        \subcaption{Power model $P(A.I.,f)$ vs measured.}
        \label{fig:iccd_fit_power}
    \end{minipage}
    \caption{Throughput and power on RTX~6000~Ada, modeled vs. measured.}
    \label{fig:iccd_model_fit_block}
    \vspace{-3mm}
\end{figure}

\noindent\textbf{Model Fitting and Extended Throughput Model.} 
Equations~\ref{eq:vanilla_throughput} and~\ref{eq:vanilla_power} are too idealized to fit the measured data. As shown in Fig.~\ref{fig:frequency_locked_roofline}, throughput (TFLOPS) still increases slightly with arithmetic intensity in the compute-bound regime. This suggests that, as the workload moves from the balanced region into the compute-bound region, the SMs do not immediately reach their saturation limit. In other words, the vanilla throughput model does not fully reflect the observed performance behavior, motivating the modified model:
\begin{equation}
    \Theta(A.I., f) = \min\left(\eta \cdot f \cdot \left(\frac{A.I.}{A.I. + \omega \cdot f^\gamma}\right), \,\, \beta \cdot A.I.\right)
    \label{eq:modified_performance_model}
\end{equation}
In Eq.~\ref{eq:modified_performance_model}, the new term in the compute-bound region represents the saturation model of arithmetic intensity. It slowly converges to 1 as A.I. increases to infinity, thus reaching the real roofline.
Using the roofline benchmark data we collected on RTX 6000 Ada GPUs, we validate the fit of the improved throughput and power model in Fig.~\ref{fig:iccd_fit_throughput} and Fig.~\ref{fig:iccd_fit_power}, respectively.
This model lays the foundation for our optimal selection of frequency based on the workload and GPU characteristics.

\begin{figure}[t]
    \centering
    \includegraphics[width=\linewidth]{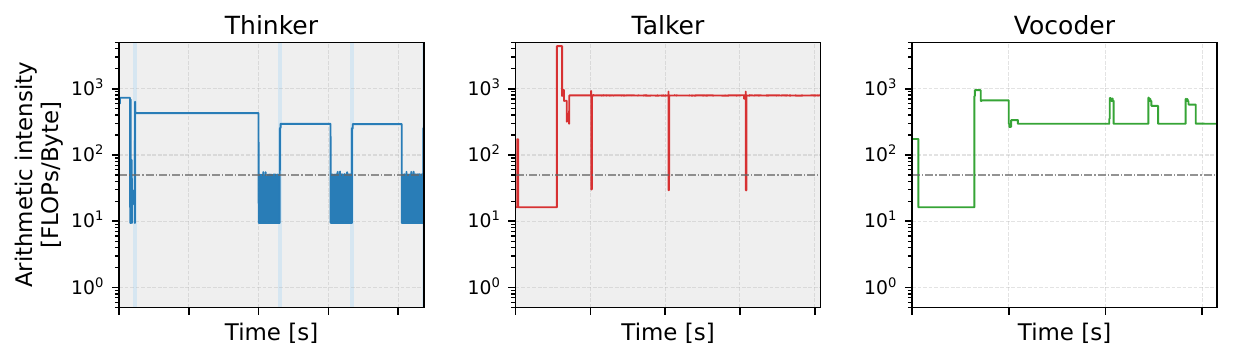}
    \caption{Arithmetic intensity of kernels by phase and stage. %
    }
    \label{fig:iccd_phase_ai_timeline}
    \vspace{-3mm}
\end{figure}

\noindent\textbf{Selecting Optimal A.I.-aware Frequency. }
At runtime, we first obtain an \textit{effective time-weighted
A.I.} of a phase, $\overline{A.I.}_{phase}$:
\begin{equation}
\overline{A.I.}_{phase} = \frac{\sum_{i=1}^{N} d_i\,A.I._i}{\sum_{i=1}^{N} d_i}
\end{equation} %
, where $d_i$ is the duration of kernel $i$, which has $A.I._i$. The individual kernel's arithmetic intensity is obtained with NVIDIA Nsight Compute (NCU). Figure~\ref{fig:iccd_phase_ai_timeline} illustrates the A.I. of each phase and stage of Qwen2.5-Omni-7B. The light band of the same color around each line is the 10th–90th percentile spread of per-kernel A.I. within each window. %

Given $\overline{A.I.}_{phase}$, the controller then aims to identify %
the lowest power that achieves the highest throughput and the highest frequency that achieves this lower power through the following optimization: 
\begin{equation}
\begin{aligned}
f^{*}_{\text{phase}} = \, & \arg\min_{f \in [f_{\min}, f_{\text{limit}}]} P\bigl(\overline{A.I.}_{\text{phase}}, f\bigr) \\
\text{s.t.} \quad & \Theta\bigl(\overline{A.I.}_{\text{phase}}, f\bigr) \geq (1-\epsilon)\,\Theta\bigl(\overline{A.I.}_{\text{phase}}, f_{\max}\bigr).
\end{aligned}
\end{equation}
, where $\epsilon$ is a performance trade-off factor. For example, $\epsilon = 5\%$ means we can tolerate a $5\%$ loss in throughput. We define $f_{limit}$ as the maximum frequency that is supported for a given A.I., as shown by the Auto-boost line in Figure~\ref{fig:frequency_locked_roofline}. The maximum supported frequency varies by A.I. due to the PMU dynamically allocating power between the GPU cores and memory. When compute and memory are balanced, significant power is allocated to memory, resulting in the GPU core frequency being throttled.

Since we need to solve for the above optimization formulation for every decode period, we require a low-overhead solver algorithm. Given our problem formulation, we could solve it using sequential quadratic programming; however, that would be relatively expensive. Since the frequency levels are discrete and monotonic with power and performance (as shown in Figure~\ref{fig:frequency_locked_roofline}), we can instead solve the above optimization formulation through binary search, enabling fast convergence to an optimal A.I.-aware frequency level.

\subsection{Thermal Headroom-aware Frequency Scaling for High Arithmetic Intensity Phases}
\label{sec:system-thermal}

To avoid the frequency degradation due to thermal throttling of compute-heavy phases, we propose a two-phase \textit{pace-and-race} frequency scaling policy that aims to \textit{pace} the SM initially by selecting a lower frequency to conserve thermal headroom, then \textit{race} by enabling auto-boost to use a higher frequency due to the conserved thermal headroom of pacing. To achieve this, we need to determine the pace and race frequency and the duration of pacing and racing. For the pace frequency, we choose the highest frequency that can run sustainably across all A.I., indicating that this frequency does not exhaust the thermal headroom, which would require the hardware PMU to intervene. As shown in Figure~\ref{fig:frequency_locked_roofline} (bottom), we select a pace frequency of $\sim$1800\,MHz. For the racing frequency, we default to PMU-guided auto-boosting to take advantage of the extra thermal headroom. We find that a pacing duration of 10\% of prefill, with 90\% racing, achieves the best balance.

\section{Evaluation}
\label{sec:evaluation}

\begin{table*}[ht]
\centering
\caption{\textbf{Ablation Study of Tri-serve Components.} Measured on Qwen2.5-Omni-7B using a 3$\times$ RTX 6000 Ada disaggregated pipeline. Energy is normalized to output tokens to reflect generative efficiency. %
}
\label{tab:ablation}
\resizebox{\textwidth}{!}{%
\begin{tabular}{@{}lccccc@{}}
\toprule
\textbf{Configuration} & \textbf{Energy / Output Tok (J)} & \textbf{Mean TTFT (ms)} & \textbf{Mean TPOT (ms)} & \textbf{Throughput (tok/s)} & \textbf{Peak Temp ($^\circ$C)} \\ \midrule
Auto-boost (Baseline)  & 4.12                            & 185                     & 48.2                    & 181.7                       & 74                             \\
Stall-aware scaling only   & 3.58 (-13.1\%)                  & 186                     & 48.3                    & 181.5 (-0.1\%)                         & 72                             \\
A.I.-aware scaling only     & 3.85 (-6.5\%)                   & 192                     & 51.5                    & 178.2 (-1.9\%)                       & 68                             \\
Thermal-aware scaling only & 4.05 (-1.7\%)                   & 188                     & 47.8                    & 184.3 (+1.4\%)              & 65                             \\ \midrule
\textbf{Tri-serve (All components)} & \textbf{3.21 (-22.1\%)}         & \textbf{190}            & \textbf{49.1}           & \textbf{183.1 (+0.8\%)}     & \textbf{64}                    \\ \bottomrule
\end{tabular}%
}
\end{table*}

\begin{table*}[t]
\centering
\footnotesize
\caption{\textbf{System Performance and Energy Efficiency Analysis.} Comparison of \textit{Tri-serve} against \textit{Auto-boost}, \textit{Fixed-frequency}, and \textit{throttLL'eM}. Energy ($E$) is Joules per output token. TTFT ($TF$) and TPOT ($TP$) report $P_{90} / P_{99}$ tail latencies in milliseconds. Offline scenarios report mean TTFT and throughput ($Thr$, tok/s). Percentages in parentheses indicate energy reduction relative to the Auto-boost baseline. Evaluated on Qwen2.5-Omni-7B with 50 prompts per run.}
\label{tab:system_comparison_transposed}
\setlength{\tabcolsep}{2.6pt}
\begin{tabular}{@{}ll cccc c cccc@{}}
\toprule
& & \multicolumn{4}{c}{\textbf{2-GPU Configuration}} & & \multicolumn{4}{c}{\textbf{3-GPU Configuration}} \\
\cmidrule(lr){3-6} \cmidrule(lr){8-11}
\textbf{Scenario} & \textbf{Metric} & \textbf{Auto-boost} & \textbf{Fixed-Med} & \textbf{throttLL'eM} & \textbf{Tri (Ours)} & & \textbf{Auto-boost} & \textbf{Fixed-Med} & \textbf{throttLL'eM} & \textbf{Tri (Ours)} \\ \midrule

\multirow{3}{*}{\textbf{Offline}} 
 & $E$ (J/tok) & 5.24 & 4.85 (-7.4\%) & 4.89 (-6.7\%) & \textbf{4.15 (-20.8\%)} & & 4.12 & 3.90 (-5.3\%) & 3.94 (-4.4\%) & \textbf{3.21 (-22.1\%)} \\
 & $TF$ (mean) & 210 & 245 & 218 & \textbf{215} & & 185 & 220 & 189 & \textbf{190} \\
 & $Thr$ (tok/s) & 142 & 128 & 138 & \textbf{141} & & 182 & 162 & 178 & \textbf{183} \\ \midrule

\multirow{3}{*}{\textbf{Online ($\lambda=0.5$)}} 
 & $E$ (J/tok) & 5.65 & 5.15 (-8.8\%) & 5.18 (-8.3\%) & \textbf{4.38 (-22.5\%)} & & 4.48 & 4.15 (-7.4\%) & 4.18 (-6.7\%) & \textbf{3.42 (-23.7\%)} \\
 & $TF$ ($P_{90/99}$) & 235 / 250 & 295 / 305 & 255 / 270 & \textbf{242 / 258} & & 185 / 190 & 230 / 238 & 195 / 205 & \textbf{191 / 195} \\
 & $TP$ ($P_{90/99}$) & 50.8 / 52.4 & 54.2 / 56.1 & 53.5 / 55.2 & \textbf{51.5 / 53.2} & & 47.9 / 49.2 & 49.5 / 51.4 & 48.8 / 50.5 & \textbf{48.8 / 50.2} \\ \midrule

\multirow{3}{*}{\textbf{Online ($\lambda=1.0$)}} 
 & $E$ (J/tok) & 5.41 & 4.95 (-8.5\%) & 4.99 (-7.8\%) & \textbf{4.22 (-22.0\%)} & & 4.25 & 4.02 (-5.4\%) & 4.05 (-4.7\%) & \textbf{3.30 (-22.4\%)} \\
 & $TF$ ($P_{90/99}$) & 240 / 255 & 302 / 310 & 265 / 285 & \textbf{248 / 262} & & 188 / 192 & 235 / 240 & 205 / 215 & \textbf{194 / 198} \\
 & $TP$ ($P_{90/99}$) & 51.2 / 53.1 & 55.4 / 58.0 & 54.2 / 56.5 & \textbf{52.1 / 54.5} & & 48.2 / 50.5 & 50.8 / 52.6 & 49.5 / 51.8 & \textbf{49.1 / 51.2} \\ \midrule

\end{tabular}
\end{table*}

\subsection{Evaluation Methodology}
\label{subsec:system_setup}
\subsubsection{Hardware Platforms}

We evaluate on a workstation with 8 NVIDIA RTX 6000
Ada GPUs with 48\,GB GDDR6 and a TDP of 300\,W.
The host infrastructure is powered by a dual-socket AMD EPYC 7543 processor configuration with a total of 64 physical cores (32 cores per socket) and Simultaneous Multithreading (SMT) disabled to ensure deterministic execution baselines. The system is provisioned with 2.0\,TiB of total system RAM to comfortably eliminate host-side memory bottlenecks during heavy continuous batching workloads.

\subsubsection{Software Stack}
Our evaluation uses an instrumented vLLM-Omni serving stack running Qwen2.5-Omni-7B with Thinker, Talker, and Code2Wav stages. The serving system is running with PyTorch 2.5.1 and Python 3.12.7. 

\subsubsection{Workloads}

We run Qwen2.5-Omni-7B on our instrumented vLLM-Omni serving engine with either a 3-GPU stage
configuration (one stage per GPU) or 2-GPU stage configuration (Thinker and Vocoder share a GPU), as indicated in the experiment. Queries
are drawn from MME-Unify~\cite{xie2025mme} (mixed-modality: text-to-audio, image-to-text,
video-to-text) and SeedTTS~\cite{anastassiou2024seed} datasets.

As previously discussed in Section~\ref{sec:motivation-ai}, we also develop a microbenchmark
compute kernel to sweep across A.I. $\in \{1, 25, 50, 100, 200\}$ FLOPs/B
and SM frequency across the supported SM-clock range. This microbenchmark runs continuously
until steady-state power and temperature are reached.

\subsubsection{Baseline Policy Definitions}
We evaluate Tri-serve against different baselines. 
\textit{Auto-boost (default)} uses NVIDIA's default auto-boost mechanism that is guided by the hardware PMU.
\textit{Fixed frequency (medium)} locks the GPU at 75\% of maximum frequency.
\textit{throttLL'eM}~\cite{10946751} is a state-of-the-art predictive frequency throttling policy for text-based LLMs. throttLL'eM is designed for unimodal LLMs; thus, in our evaluation, we apply throttLL'eM only to the prefill and decode phases of Thinker and Talker.

\subsubsection{Metrics}
We evaluate Tri-serve and the baselines on several metrics of interest. We measure throughput as tokens per second, service quality as time-to-first-token (TTFT) and time-per-output-token (TPOT). Energy efficiency is evaluated as total energy in Joules over output tokens to obtain Joules per output token.

\subsection{Ablation Results}
\label{sec:eval-ablation}
\label{sec:eval-summary}
We evaluate Qwen2.5-Omni-7B on three NVIDIA RTX 6000 Ada GPUs, with one stage per GPU. Table~\ref{tab:ablation} shows the ablation results of how each component benefits multimodal inference energy efficiency.

\noindent\textbf{\textit{Stall-aware idle frequency scaling} only.} 
Our stall-aware idle scaling policy accurately identifies dependency stalls and reduces the stall frequency to
210\,MHz for SMs and 810\,MHz for memory, yielding a
\textbf{-13.1\%} reduction in energy per output token with no change to
TTFT, TPOT, or throughput.

\noindent\textbf{\textit{A.I.-aware frequency scaling} only. } 
Our A.I.-aware frequency scaling policy selectively reduces the frequency of memory-bound decode phases by selecting a frequency level that achieves the lowest power at baseline throughput levels. 
The A.I.-aware frequency policy achieves a \textbf{-6.5\%} reduction in energy per output token with minor impact on TTFT, TPOT, and throughput. Due to the longer decode phases, we also see a more drastic reduction in peak GPU temperature, from 74$^\circ$C in the baseline to 68$^\circ$C with this technique.

\noindent\textbf{\textit{Thermal-aware frequency scaling} only. } 
Our Thermal-aware \textit{pacing} mechanism selects a sustainable frequency level that conserves thermal headroom to enable a higher frequency auto-boosting \textit{racing} period to achieve better performance during compute-intensive periods. This achieves a \textbf{-1.7\%} improvement in energy per output token, with \textbf{1.4\%} higher throughput, and a peak temperature of 65$^\circ$C.

\noindent\textbf{Full Tri-serve. } Combined, Tri-serve minimizes unnecessarily high frequency during dependency stalls and low A.I. decode periods, and avoids thermal throttling during compute-heavy prefill. Compared to the hardware PMU-managed auto-boost baseline, Tri-serve achieves a \textbf{-22.1\%} reduction in energy per output token, with a \textbf{0.8\%} improvement in throughput, while having the lowest peak temperature, only 64$^\circ$C. This demonstrates that, with thermal-aware control, GPUs can run cooler, at lower power, and with sustained performance.

\subsection{Baseline Comparison Results}
\label{sec:baseline_comparison}

In Table~\ref{tab:system_comparison_transposed}, we evaluate Tri-serve against a baseline of hardware PMU-guided auto-boosting and a fixed frequency of 1500\,MHz that can run sustainably without thermal throttling.
We also evaluate against an implementation of SOTA throttLL'eM~\cite{10946751} which aims to select the lowest power that satisfies SLO. We evaluate both a 3-GPU scenario (as mapped in Figure~\ref{fig:vllm_omni_pipeline}) and a 2-GPU scenario where the Thinker and Vocoder share a single GPU. We test under one offline scenario and two online scenarios with varying request arrival rates ($\lambda$) of 0.5 and 1 RPS.
In all scenarios, a fixed frequency achieves 5.3\%--8.8\% better energy efficiency than auto-boost, at the cost of higher TTFT and TPOT.
Because throttLL'eM was designed for unimodal LLMs and has limited coverage of multimodal model pipelines, its energy savings are limited to a 4.4\%--8.3\% improvement, but with better TTFT and TPOT than the fixed-frequency scenario. %
Tri-serve universally achieves the best energy efficiency, with a 20.8\%--23.7\% improvement, while nearly matching the baseline auto-boost TTFT and TPOT metrics. This demonstrates Tri-serve's benefits across a range of baseline scenarios, showing that it is a low-complexity yet effective strategy for energy-efficient multimodal serving.

\section{Related Work}
\label{sec:related}

\textit{Multimodal and large-model serving.}
Prior serving systems focus on throughput and latency for continuous LLM requests. Orca~\cite{280922}, vLLM~\cite{10.1145/3600006.3613165}, Sarathi~\cite{298679}, AlpaServe~\cite{288582}, and DistServe~\cite{zhong2024distserve} improve batching, memory management, or stage disaggregation for text-only serving, while vLLM-Omni~\cite{yin2026vllmomni} extends disaggregated serving to multimodal pipelines. In contrast, Tri-serve exploits dependency stalls, arithmetic intensity, and thermal headroom as control inputs for DVFS.

\textit{Power and energy-aware inference.}
Energy-aware unimodal LLM inference methods include POLCA~\cite{10.1145/3620666.3651329}, \mbox{throttLL'eM}~\cite{10946751}, $\mu$-Serve~\cite{298496}, and PowerInfer~\cite{10.1145/3694715.3695964}. Related characterization work studies cluster-level energy and power behavior~\cite{tabbakh2024towards,jia2024pccl,9195730,10.1145/3634769.3634807}, while execution-idle behavior in GPU clusters has also been observed~\cite{lei2026energy}. These systems typically optimize a single phase or treat thermal effects reactively; Tri-serve instead combines stall-aware, A.I.-aware, and thermal-aware control in one controller.

\textit{Thermal modeling and pacing.}
Our thermal model follows classic thermal and leakage-power work~\cite{1650228,5450539,10.1145/1815961.1815998}, and the roofline-based A.I. model follows prior roofline studies~\cite{williams2009roofline,nugteren2014roofline,guerreiro2018gpgpu}. Pace-then-race controllers have been used in other systems~\cite{uDPM,zhou2020-swan}, but we specialize to multimodal prefill and integrate with A.I.- and stall-aware scaling.

\section{Conclusion}\label{sec:conclusion}

We identified three sources of GPU power inefficiency in modern multimodal serving: \textbf{(1)} dependency stalls that leave SM and memory clocks near boost levels while the GPU makes no progress; \textbf{(2)} an anti-correlation between auto-boost frequency and arithmetic intensity, which raises frequency in low-intensity phases and lowers it in compute-intensive ones; and \textbf{(3)} thermal throttling that steadily reduces SM frequency during compute-heavy prefill.
To address these inefficiencies, we introduced Tri-serve, a software-level DVFS controller that combines stall-aware idle scaling, A.I.-aware decode locking, and thermal-aware pace-and-race scaling.
Across Qwen2.5-Omni-7B workloads, Tri-serve reduces energy per output token by 20\%--23\% relative to auto-boost, while keeping throughput within 3\% and TPOT within 2\%.

\section*{Acknowledgment}
We thank the anonymous reviewers for their invaluable comments and suggestions. This research was supported in part by the University of California, Riverside; the National Science Foundation under grants CNS-2415202, CNS-2047521, CCF-2324940, CCF-2324941, and NSF-MeitY Grant (e-file: 3149156); the ANRF Grant (ANRF/ECRG/2024/006088/ENS); the MeitY Grant (VARCOE/23/02); and IITBBS Project RP433.

\bibliographystyle{IEEEtran}

\bibliography{refs}

\end{document}